\documentclass[letterpaper]{article} %
\usepackage[draft]{aaai2026}  %
\usepackage{times}  %
\usepackage{helvet}  %
\usepackage{courier}  %
\usepackage[hyphens]{url}  %
\usepackage{graphicx} %
\usepackage{natbib}  %
\usepackage{booktabs}

\usepackage{caption} %
\usepackage{algorithm}
\usepackage{algorithmic}
\usepackage{xcolor}
\newcommand{\answerYes}[1]{\textcolor{blue}{#1}} 
 
\newcommand{\answerNA}[1]{\textcolor{gray}{#1}} 
 
\usepackage{array}
\usepackage{enumitem}

\newif\ifcomment
\commenttrue
\ifcomment
\newcommand{\sz}[1]{{\textcolor{red}{\textbf{SZ: #1}}}}
\else
\newcommand{\sz}[1]{}
\fi

\usepackage{newfloat}
\usepackage{listings}
\DeclareCaptionStyle{ruled}{labelfont=normalfont,labelsep=colon,strut=off} %
\floatstyle{ruled}
\newfloat{listing}{tb}{lst}{}
\floatname{listing}{Listing}

\copyrighttext{Accepted for publication at the 21st International AAAI Conference on Web and Social Media (ICWSM 2027). Please cite accordingly.}

\title{``@Grok Put Her in a Bunny Suit:'' An Empirical Analysis of Non-Consensual Synthetic Intimate Imagery on X}
\author{
    Nicolò Fontana\equalcontrib,
    Vahid Rahimzadeh\equalcontrib,
    Savvas Zannettou
}
\affiliations{
    Delft University of Technology\\
    \{N.Fontana,V.Rahimzadeh, S.Zannettou\}@tudelft.nl
}

\begin{document}

\maketitle

\begin{abstract}

Conversational AI systems are increasingly embedded into social media platforms, where model outputs can be generated and distributed in public settings. This integration creates new opportunities for abuse because users can request, produce, and circulate harmful content within the same platform-native interaction. One particularly concerning form of abuse is non-consensual synthetic intimate imagery (NSII): AI-generated sexualized depictions of real people produced or shared without their consent. Despite growing public concern, little is known about how NSII unfolds in public, platform-integrated AI systems, including its prevalence, the forms it takes, and who it targets.
To shed light on these aspects, we study Grok on X. We collect 2M public Grok conversations comprising 5.9M tweets between February 19 and March 27, 2026, and focus on 50K conversations in which Grok generated an image. Combining an open-source multimodal model (Qwen3.5) with human annotation, 
we estimate the prevalence of NSII and find an estimate between 2.94\% and 3.05\% of analyzed conversations. We then use concept induction and target-level annotation to characterize how this abuse manifests and who it targets.
We find that Grok-enabled NSII is recurring, diverse, and targeted. It often operates through incremental image-editing requests, including changes to clothing, exposure, pose, body appearance, and scene context. We also find that targeting is highly individualized and disproportionately directed at feminine-presenting people. Although many targets are public-facing figures such as celebrities, content creators, and politicians, private individuals account for nearly one quarter of target entries. Our work shows that public, platform-integrated conversational AI systems can create systemic risks that require safeguards at the conversation level, effective reporting mechanisms, and independent auditing.

\textbf{\textcolor{red}{Disclaimer: This paper discusses NSII and may be disturbing to some readers.}}
\end{abstract}

\section{Introduction}

Conversational AI systems are no longer confined to private chat windows.
They are increasingly being embedded into public online platforms, where they can participate directly in social interaction. Grok is a prominent example of this shift. On X, users can publicly invoke Grok in a conversational thread, and its responses appear there. This integration has clear, legitimate uses; recent work by~\citet{mei2026grok} shows that people use Grok to seek general information, fact-check, and engage in debate.
At the same time, embedding a conversational AI system directly into a public interaction space creates new opportunities for abuse, especially when limited guardrails are in place.

One important form of abuse in this setting is non-consensual synthetic intimate imagery (NSII)~\cite{umbach2024non}: AI-generated sexualized depictions of an individual that are produced or shared without their consent. NSII falls within the broader landscape of image-based sexual abuse~\cite{mcglynn2017image,henry2019image} and can cause serious harm to victims, such as significant psychological distress~\cite{bates2017revenge,mcglynn2021s}.
In the weeks following xAI's rollout of Grok's image-editing capabilities in December 2025, researchers at the Center for Countering Digital Hate estimated that millions of sexualized images of real people were generated on X, including depictions of minors~\cite{ccdh-2026}.
This phenomenon has already drawn regulatory scrutiny: in early 2026, the EU Commission opened a formal investigation against X under the Digital Services Act~\cite{EU_DSA_2022}, citing risks of manipulated sexually explicit imagery and potential child sexual abuse stemming from Grok's integration into the platform~\cite {commision-investigation}.
What makes this case especially important is that Grok’s integration into X does not merely enable NSII; it can also compress the process of requesting, generating, and publicly circulating such content into a single platform-native interaction.

Historically, producing NSII of a real person involved a four-step pipeline: (1) acquiring a capable generation tool, typically a dedicated nudifier app, website, or fine-tuned model; 
(2) obtaining and uploading source imagery of the target;
(3) generating the synthetic output in a private environment; 
and (4) separately distributing the output on a public online channel. 
Each of these steps imposed a cost: a technical barrier, a separate hosting surface, and, critically, a gap between production and distribution. 
Publicly integrated conversational AI systems like Grok on X collapse this pipeline into one step. The tool lives inside the platform. Source imagery is either already on the platform or attached to the invoking reply. Generation is itself a public post. 
So production, distribution, and targeting are no longer separate acts; they are one.
Because this setting differs structurally from earlier ones, existing evidence does not yet tell us how this abuse unfolds in practice on a public, platform-integrated system like Grok on X.

Despite growing public concern, we still know little about how this abuse manifests in practice on platform-integrated conversational AI systems. Most of what is publicly known about this phenomenon comes from journalistic reporting~\cite{bloomberg-grok,reuters-grok} and audits from non-profit organizations~\cite{ccdh-2026}. These sources have been essential in raising the alarm, but they leave core empirical questions unanswered. 
Meanwhile, the academic literature on image-based sexual abuse has studied distinct settings: dedicated deepfake-pornography sites~\cite{han2025characterizing,timmerman2023studying}, standalone nudifier applications~\cite{williams2025there, gibson2025analyzing}, and the misuse of general-purpose generative models in closed or private contexts~\cite{qu2023unsafe,li2024safegen,hawkins2025deepfakes}.
Related work has also examined reporting mechanisms to mitigate the spread of such content online~\cite{qiwei2024reporting,qiwei2026platforms}.
Each of these settings presupposes the four-step pipeline described above, in which generation and distribution are separate acts on separate surfaces. The setting we study, in which that pipeline has collapsed into a single platform-native act performed in public, is, to the best of our knowledge, empirically unexamined by the research community.
As a result, we lack a systematic measurement of how prevalent this behavior is, what forms it takes, and who it targets.
Specifically, this study asks:
\begin{itemize}[noitemsep, topsep=0pt]
    \item \noindent \textbf{RQ1: Prevalence.} How prevalent is the use of Grok on X to generate NSII?
    \item \noindent \textbf{RQ2: Manifestation.} What types of NSII requests resulted in the generation of NSII content from Grok?
    \item \noindent \textbf{RQ3: Targets.} Who is targeted by Grok-enabled NSII?
\end{itemize}

To answer these questions, we design a data collection methodology to obtain a random sample of Grok interactions on X. Using X's Web API, we collect 2M conversations comprising 5.9M tweets between February 19, 2026, and March 27, 2026, and then focus on approximately 50K conversations in which Grok generated an image. For RQ1, we combine Qwen3.5, an open-source vision-language model, with human annotation to identify NSII, yielding 1.5K conversations containing such content. For RQ2, we summarize these conversations with the same model and use the LLooM framework~\cite{lam2024conceptInduction} to identify recurring request patterns. For RQ3, we perform annotation with Qwen3.5 to understand who is targeted.

Our study makes the following main findings:
\begin{itemize}[noitemsep, topsep=0pt]
    \item \textbf{RQ1: Prevalence.} NSII generation is a non-negligible component of Grok's image-generation behavior on X. We find that between 2.94\% and 3.05\% of image-generation conversations in our dataset ($\approx$1.5K out of 50K) result in an NSII output. Also, we find that the generation of NSII content is persistent across the entire observation period of our study, highlighting the potential that this risk is systemic.
    \item \textbf{RQ2: Types of NSII.} Grok-powered NSII requests are diverse and take the form of incremental image edits that sexualize a person. The most common request types involve increasing revealingness (60.4\%), changing the target into a sexualized outfit (54.1\%), and removing clothing or increasing nudity (42.2\%). Users also request more specific transformations, including costume swaps (28.4\%), added visual effects (25.5\%), pose or angle changes (20.2\%), material or transparency edits (14.7\%), body-part enlargement (10.3\%), garment-structure edits (9.9\%), swimwear replacement (5.2\%), and facial-feature modification (5.1\%). These findings show that Grok-enabled NSII is not limited to explicit nudification requests; it spans a broader set of editing workflows. %
    \item \textbf{RQ3: Targets of NSII.} Our analysis shows that Grok-generated NSII is primarily individualized, strongly gendered, and not limited to celebrities. Most NSII-positive conversations targeted a single person (85.9\%), and most target entries were feminine-presenting (80.9\%). At the same time, targets spanned content creators (25.1\%), celebrities (23.9\%), private individuals (22.7\%), adult content creators, politicians, and other public-facing figures. The most frequently identified named targets further show the global character of this abuse, spanning entertainment, politics, technology, and media figures across multiple countries. These findings reinforce that Grok-enabled NSII is a systemic risk and not only a narrow abuse problem that affects celebrities.

\end{itemize}

\section{Background \& Related Work}

\subsection{Grok}

Grok is a conversational AI system that was introduced by xAI in 2023~\cite{grok-intro}, and since December 2024, it has been integrated into the X platform~\cite{grok-intro-x}.
Nowadays, on X, users can publicly invoke Grok inside conversational threads, and Grok's responses appear in the same thread.
This significantly changes the setting in which conversational AI systems are used. 
Whereas systems such as OpenAI’s ChatGPT are typically accessed through private one-to-one interfaces~\cite{karnam2026bowling}, Grok also operates in a public, social, and highly visible environment.
Characterizing this public setting is essential, as model use, user intent, and the dissemination of AI-generated content can all occur within the same interaction.

Prior work has begun to characterize how Grok is used in this public setting.
\citet{mei2026grok} provide a broad empirical analysis of how Grok is used publicly on X. 
By collecting and analyzing 41K public Grok interactions, they show that users most often use Grok for information seeking, fact-checking, and engaging in debates. At the same time, they show that users assign different roles to Grok, like truth arbiter, advocate, and adversary. Other studies focus more specifically on Grok's role in verifying information and correcting misinformation.
\citet{RenaultTea25p_GrokThisTrue} analyze 1.6M fact-checking requests to Grok and Perplexity on X, showing that the majority of requests were related to politics.
Also, the study shows significant agreement among Grok, Perplexity, and Community Notes for fact-checking.
\citet{caramancion2026using} examines Grok's misinformation correction and finds that corrections involving Grok are associated with fewer ad hominem attacks than direct human-issued corrections. 
Altogether, these studies show that Grok is increasingly used for verifying claims and can have an important role in public discourse on X.

While prior work has examined the general use of Grok on X, the research community still lacks a systematic understanding of the harms that may arise from Grok’s public integration into the platform. Journalistic reporting provides early evidence of these risks, including toxic and antisemitic outputs~\cite{grok-toxic} and the generation of sexual deepfakes or nude images~\cite{guardian-nudification}. More systematic evaluations remain limited. Researchers at the Center for Countering Digital Hate estimate that, after the rollout of Grok's image-editing feature, Grok generated millions of sexualized images on X, including images depicting children~\cite{ccdh-2026}. The closest prior research is the audit by \citet{lefkowitz2026grok}, which analyzes 43 incidents of targeted sexualization involving Grok-generated images. The audit identifies proxy terms that users employed to evade safeguards and argues that moving image generation behind a paid tier monetized the feature without addressing the underlying risk. Overall, these reports and audits show that Grok can be used for harmful public image generation; however, they do not provide a large-scale empirical characterization of the phenomenon. As a result, we still know little about how prevalent Grok-enabled NSII is, how it manifests, and who the targets are.

\subsection{NSII}

The term NSII~\cite{umbach2024non} refers to AI-generated depictions of a real person that are produced or shared without consent. NSII is a form of image-based sexual abuse (IBSA), which broadly refers to the non-consensual creation or distribution of nude or sexual images~\cite{mcglynn2017image}. Prior work has documented the serious harms experienced by victims of IBSA. Drawing on interviews with 75 participants, \citet{mcglynn2021s} show that these harms extend beyond immediate distress and include social rupture, isolation, and existential threat. These findings underscore the urgency of understanding and mitigating IBSA. At the same time, generative AI may amplify this problem by lowering the cost and technical barriers required to create sexualized imagery of real people. As a result, recent work has begun to examine NSII as an emerging AI-mediated form of IBSA.

One line of this work focuses on the dedicated tools, communities, and marketplaces that support NSII creation. \citet {gibson2025analyzing} presents a systematic study of 20 popular AI nudification websites and examines how they position themselves, what features they offer, and how they monetize their services. They find that many websites target women and offer diverse harmful capabilities like undressing and rendering individuals in sexual positions. Similarly,~\citet{williams2025there} analyzes 29 undressing apps and finds that these services facilitate and encourage the creation of NSII through their platform affordances, marketing, and revenue models. 
Prior work has also studied dedicated deepfake pornography communities and marketplaces.
~\citet{timmerman2023studying} examines the /r/deepfakes subreddit and the MrDeepFakes forum, showing how these online communities combine technical discussion, community-building, and hosting of deepfake content.
More recently,~\citet{han2025characterizing} characterized the Mr. DeepFakes forum to analyze the marketplace's scale, target population, paid request economy, and user discussions about deepfake creation.
~\citet{cui2026celebrities} perform an empirical analysis of such content on 4chan, aiming to characterize the models used by 4chan users to generate NSII images, as well as the dynamics between the requesters and providers of such content.

Others examine how generative models can produce sexual or unsafe imagery, and how such behavior can be mitigated.
~\citet{qu2023unsafe} study unsafe image generation in text-to-image models, showing that they can generate sexually explicit, violent, hateful, and other unsafe content.
\citet{hawkins2025deepfakes} examine the accessibility of deepfake models on platforms such as Huggingface and Civitai, finding that many models are available for generating non-consensual depictions of people.
Complementing these studies,~\citet{li2024safegen} propose SafeGen, a mitigation strategy designed to reduce the generation of sexually explicit images by removing unsafe visual representations from text-to-image models.

Recent work has also examined public perceptions of NSII and the reporting systems available to victims. \citet{brigham2024violation} survey U.S. participants about AI-generated non-consensual intimate imagery and find strong opposition to its creation and sharing, especially when the content depicts sexual acts, while also showing that attitudes vary by gender, relationship to the creator, and views on sexual consent. Complementing this perception-focused work, \citet{qiwei2026platforms} study how victim-survivors report non-consensual intimate imagery online, finding that platform reporting processes are often opaque, hostile, and ineffective. \citet{qiwei2024reporting} further demonstrates these failures through an audit study on X, showing that reports about non-consensual sexual content resulted in no removals over three weeks, while copyright-based reports led to fast removals.

Finally,~\citet{cuevas2026deepfake} investigate the effect of NSII-specific regulation on three different forums used for generating NSII, finding that the introduction of a new regulation in the US did not suppress NSII-related activity in the investigated forums.

\smallskip \noindent \textbf{Remarks.} Prior work focused primarily on dedicated nudification tools, deepfake communities, and model-level risks, leaving less understood how NSII unfolds when AI models are embedded directly into a public social platform. Our work addresses this gap by studying Grok on X, measuring the prevalence of NSII, and characterizing its manifestation.

\section{Data Collection}

We collected publicly visible Grok interactions on X by repeatedly monitoring Grok's replies page (\url{https://x.com/grok/with_replies}) throughout the day. Our goal was to obtain an approximately random sample of recent public Grok-user conversations, rather than a complete historical archive. Because new interactions appear continuously, we designed the collection process to repeatedly sample the most recent Grok exchanges visible at randomly distributed points in time.
To implement this strategy, we used ten sockpuppet accounts, that is, measurement accounts created solely for data collection. We used multiple accounts because the web API endpoint underlying the replies page requires authentication, and consolidating all requests into a single account would have increased the likelihood of that account being rate-limited or blocked. Each account visited Grok's replies page at random intervals each day and retrieved the most recent Grok interactions visible at that moment. In practice, each retrieval typically returned around twenty recent conversations. 
This approach helped approximate a rolling random sample of Grok's recent public activity.

Rather than scraping, each sockpuppet account queried the same  API endpoint used by X's Web client to power the replies page. Specifically, the collection relied on the public website endpoint exposed via X's Web interface, namely \emph{UserTweetsAndReplies}, along with the query parameters required to retrieve Grok's recent reply conversations. 
The endpoint returns each conversation in a structured JSON format that includes author and other metadata for every post. These metadata allowed us to distinguish Grok-generated images from user-uploaded images. 
Also, this approach allowed us to observe the same public interactions that an ordinary user would see when visiting Grok's replies page through the platform interface.
Our method was intentionally designed to be observational and minimally disruptive. We only collected publicly visible interactions, limited ourselves to the platform pathway already used by the web client, and imposed a strict limit across all accounts. We capped the total volume at approximately 3K requests per day to avoid overloading the platform or disrupting normal operations. Under this budget, our data collection pipeline collected approximately 60K conversations per day.

We collected data continuously from February 19, 2026, to March 27, 2026, and our dataset comprises over 2,039,067 conversations, totaling 5,938,326 tweets. Since our goal is to analyze how Grok is used to generate NSII, we filter our dataset to include only conversations in which Grok generates an image. We do this by filtering the media URLs in the JSON data we obtain directly from X. By performing this filtering step on our dataset, we select 51,477 conversations out of 2,039,067 total.

\noindent \textbf{Data Representation.} Each collected data entry corresponds to a Grok conversation with a user and any immediately relevant surrounding context. For our analysis, the collected data is represented at the conversation level rather than as isolated, standalone tweets. This is important because harmful interactions with Grok cannot always be inferred from Grok's reply alone; they often depend on the relationship between earlier context, the user's invocation, and Grok's response. To support this type of analysis, conversations are normalized into three components:
1)~\emph{Context tweet}, which provide earlier conversational context when available;
2)~\emph{Grok invocation tweet}, which captures the user's request to the Grok model; and
3)~\emph{Grok response tweet}, which contains Grok's public reply.

\section{Methodology}

This section describes our methodology for detecting and analyzing Grok-enabled NSII on X. In a nutshell, we use open-source multimodal models, along with human annotations, to detect NSII and characterize its types and targets. %

\subsection{Detecting NSII}

\noindent \textbf{Automated Detection.} To identify conversations where Grok is likely involved in the generation of NSII, we use a large multimodal language model, Qwen3.5-122B-A10B, which we run locally in our infrastructure. This model is designed to process multimodal input, including text and images, making it well-suited for analyzing Grok conversations that may involve image manipulation or generation. We selected this model because it was one of the most powerful open-source multimodal language models available at the time of our analysis, and it has demonstrated strong performance across a variety of tasks~\cite{qwen3_5_huggingface}.
To perform the annotation, we design a prompt that instructs the model to analyze each conversation and determine whether the user's invocation tweet explicitly requests Grok to produce NSII (\emph{NSII request label}), such as nudification or explicit content of a real person. The model is also asked to assess whether Grok's response appears to comply with that request by visually comparing the input and output images (\emph{Grok compliance label}). The model outputs binary labels for whether the invocation tweet requests the generation of an NSII imagery and another label on whether Grok complies with the user request. In addition, the model also outputs a free-text explanation of its reasoning, as well as a summary of the conversation. The full prompt is provided in the online repository~\cite{osf_repo}. 

We run the model on all 51,477 conversations that include media URLs, and we obtain the results for both the NSII request and Grok compliance labels. Note that we managed to obtain results for 50,041 out of 51,477 conversations, as the model failed to retrieve the images from X for 1,436 conversations. This is likely due to the fact that some of the media URLs included in the conversations have become inaccessible, either because the original media was deleted by the user or because of moderation actions taken by X.
The automated detection step resulted in the identification of 2,643 conversations where the model determined that the user explicitly requested Grok to produce NSII and Grok complied with the request (i.e., both binary labels were positive).

\noindent \textbf{Validation of Automated Detection.}
To validate the performance of our automated detection using the Qwen3.5 model, we extract a random sample of 200 conversations: 100 labeled as NSII and 100 labeled as non-NSII (as labeled by the Qwen3.5 model).
Then, three authors of this study independently annotated the 200 conversations, following the same instructions provided to the Qwen3.5 model.
We find a substantial inter-annotator agreement. Specifically, we calculate the Fleiss’ kappa score across the three annotators and we obtain $\kappa=0.725$ for the NSII request label and $\kappa=0.736$ for the Grok compliance label. 

Examining the disagreements between the annotators revealed several sources of uncertainty.
First, missing images from the context tweet or limited conversational context resulted in disagreements.
Also, there was a difficulty determining whether an image depicted a real person or whether the input and output depicted the same individual.
Moreover, the annotators had differing judgments about whether the imagery was intimate or whether the requested modification made the image more explicit or intimate.
These disagreements indicate that even binary judgments can depend on incomplete evidence and subjective interpretation. Accordingly, the observed inter-annotator agreement does not establish that a single annotator's judgments represent those of the other annotators.

To assess the performance of Qwen3.5 in this task, we   create the ground truth label for each conversation using the majority vote. We compare the Qwen3.5 predictions with our human-labeled validation set to understand how reliable the automated screening step is for our analysis. As shown in Table~\ref{tab:qwen-performance}, the model exhibits a high recall and lower precision for both labels: it misses very few true NSII requests or compliant Grok responses, but it also flags some non-NSII conversations as positive. This behavior is useful for our pipeline, where the model serves as a sensitive first-pass detector and human review reduces false positives. Our goal is therefore not to identify the best possible classifier for this task, but to use an LLM-assisted workflow to characterize the scale and manifestation of Grok-enabled NSII on X.

\begin{table}[t!]
\centering
\resizebox{\columnwidth}{!}{
\begin{tabular}{lrrrrrrr}
\toprule
Label & Acc. & Prec. & Rec. & F1 & FPR & FNR \\
\midrule
NSII Request & 0.825 & 0.660 & 0.985 & 0.790 & 0.255 & 0.015 \\
Grok compliance & 0.790 & 0.590 & 0.983 & 0.737 & 0.292 & 0.016 \\
\bottomrule
\end{tabular}
}
\vspace{-2mm}
\caption{Performance of the Qwen3.5 model in identifying whether the user requests NSII and Grok's compliance. We report the accuracy, precision, recall, F1 score, as well as the false positive and false negative rate of the model.}
\label{tab:qwen-performance}
\vspace{-2mm}

\end{table}

\noindent \textbf{Manual Human Annotation.} 
Given that our automated detection method yields a significant number of false positives, we aim to reduce them using manual annotations. Specifically, all 2,643 conversations labeled as positive for both NSII request and compliance by the model were manually annotated by a single annotator to confirm that they are indeed NSII. We annotate this large dataset with only one annotator because we aim to reduce the potential harm for the annotators who are exposed to disturbing content.
The human annotation process involved reviewing each conversation and verifying whether the user's invocation tweet explicitly requested the generation of NSII and whether Grok's response complied with that request, indicating that Grok was actually used to generate NSII. The human annotator was provided with the same information as the model, including the conversation text and, when available, the input and output images. The human annotation process identified 1,526 conversations in which the user explicitly requested NSII, and Grok complied with that request.

Note that determining whether these requests are genuinely non-consensual presents a challenge. We can not verify the identity of the requesting user or determine whether they are the person depicted in the input image. Consequently, we cannot directly establish whether the depicted individual consented to the generation of the content. To assess the likely non-consensual nature of these requests, a single annotator reviewed all 1,526 conversations based on how the user framed their relationship with the depicted individual. Specifically, we distinguished between first-person requests, in which users presented themselves as the person depicted, and third-person requests, in which users referred to the depicted individual as someone else. We identified 57 first-person requests and 1,469 third-person requests. Thus, the vast majority of the requests likely involved generating intimate imagery depicting another person and were therefore non-consensual. Nevertheless, linguistic framing provides only a proxy: third-person phrasing does not definitively establish the absence of consent, while first-person phrasing does not allow us to verify that the user is actually the depicted individual. We therefore cannot conclusively determine the consent status of individual instances. Given this limitation, for the remainder of the paper, we analyze the complete set of 1,526 conversations that met our request and compliance criteria.

\subsection{Detecting Types of NSII}

To identify the different types of NSII, we use LLooM~\cite{lam2024conceptInduction}, a framework that uses the power of LLMs to generate interpretable concepts from text data.
The LLooM framework starts with a \emph{seed} to focus the analysis (i.e., a user-defined LLM prompt), then \emph{distills} unstructured text by filtering relevant excerpts and summarizing them using LLM prompts. 
Then, the framework summarizes items and clusters them using embeddings and a clustering algorithm to group related ideas. An LLM then \emph{synthesizes} each cluster into human-readable concepts.
The framework can \emph{loop} over those concepts, repeating all previous steps to produce higher-level concepts.
Finally, LLooM uses a scoring procedure in which the input is scored against the concepts on a 5-point Likert scale (strongly agree to strongly disagree).

In our work, we use LLooM to characterize the different types of NSII requests in our dataset. Specifically, we instantiate LLooM with the seed prompt: \emph{``What requests do users make to Grok for image generation or modification? Focus on the type of request and not the target.''} and we provide as input the textual summaries of the 1,526 conversations where Grok generated NSII. 
This process resulted in an initial list of 12 concepts. Following LLooM’s iterative refinement workflow, we reviewed the concepts and removed one, ``Non-Consensual/Forced Change,'' because it was overly broad: rather than distinguishing a specific request pattern, it largely captured the non-consensual nature of the dataset. We therefore retain 11 concepts, reported in Table~\ref{tab:rq2_concept_set}.
Then, we use the LLooM framework to score all 1,526 conversations, and we treat a concept as appearing in a conversation if the LLooM framework scores it as ``Strongly Agree'' or ``Agree,'' following previous work that used LLooM~\cite{kirsten2025characterizing}. We provide more details on the models and parameters used in Table~\ref{tab:lloom_hyperparams} in the Appendix.

\subsection{Identifying Targets}

To identify who the targets of NSII are on X, we use the Qwen3.5 model.
We take into account the whole conversation context (text and images), and we instruct the model to annotate conversations across four dimensions:
1)~Number of NSII targets per conversation;
2)~The perceived gender representation (feminine, masculine, androgynous) of each target;
3)~The public role (e.g., Celebrity, Politician, Content Creator, Private Individual) for each target;
4)~Their name, in case they are public figures.
We provide the full prompt in the online repository~\cite{osf_repo}.

\smallskip \noindent \textbf{Validation of Target Identification.} 
To validate the performance of the Qwen 3.5 model across the four dimensions mentioned above, we extract two random samples: 200 randomly selected conversations to validate the number of targets, and 200 randomly selected target entries to validate the model's ability to identify the actual targets. 
The samples were independently annotated by two annotators without access to the Qwen3.5 annotations, except for the identified names, which were provided to assess name correctness. 
Inter-annotator agreement, measured using Fleiss' $\kappa$, was 0.661 for target count, 0.913 for perceived gender representation, and 0.441 for public role. 
The lower agreement for public role reflects the subjectivity and uncertainty involved in assessing individuals' public status from images and limited X profile information, compounded by the larger set of categories and potentially overlapping roles.
The most frequent disagreement pairs between the annotators were Private Individual and Adult Creator, Private Individual and Content Creator, and Content Creator and Adult Creator.
This pattern suggests that human annotators have difficulty distinguishing these roles consistently from the available information, warranting caution when interpreting comparisons involving these categories.
For name correctness, Fleiss' $\kappa$ was 0.530, with 90.6\% raw agreement. The predominance of positive correctness judgments increased expected chance agreement, resulting in a lower $\kappa$ despite high raw agreement. 

After discussing and resolving the disagreements, we find that the model achieved strong performance in identifying the number of targets and their perceived gender representation, with accuracies of 96\% and 99.5\%, respectively. 
For the public role, the model performed sub-optimally with an accuracy of 52\%; most of the disagreements were between the categories Private Individual, Content Creator, and Other. Most misclassifications from the Qwen model were between the categories Private individuals, content creators, and other.
Overall, determining an individual's public role often requires external information, such as web searches to establish their occupation or public presence, which was unavailable to the model in our setup. This limited context may partly explain the lower performance on this task.
Regarding the identified name, we find an accuracy of 84.9\%.

\section{Results}

\subsection{RQ1: Prevalence of NSII on X}

We first estimate how often public Grok image-generation conversations on X result in NSII. As described above, our image-generation subset contains 51,477 conversations, of which 50,041 could be processed because the relevant media remained retrievable at annotation time. Within this analyzable subset, the automated detector initially flagged 2,643 conversations as positive for both the NSII-request and Grok-compliance labels. After manual review of all flagged conversations, 1,526 were confirmed as likely cases in which a user explicitly requested NSII, and Grok produced a compliant output, corresponding to 3.05\% of analyzable public Grok image-generation conversations in our dataset.
Note that of these 1,526 conversations, 57 contained first-person requests, in which users presented themselves as the depicted individual. These requests may involve consensual generation, although we cannot verify the requester's identity. Excluding these cases from the positive count leaves 1,469 conversations involving third-person requests, corresponding to 2.94\% of the same 50,041 analyzable conversations. 
Overall, these rates show that NSII generation is not an isolated edge case in the observed public use of Grok. While 2.94-3\% may appear small in relative terms, it becomes substantial at Grok’s scale: a system that generates millions of public posts can produce a large absolute number of harmful outputs even when the misuse rate is only a few percent. This is particularly concerning because Grok is integrated directly into X.
At the same time, our prevalence estimates should be interpreted in light of our measurement pipeline. We exclude 1,436 conversations whose media could no longer be retrieved, so our estimate applies to the 50,041 retrievable image-generation conversations. Also, the final count is based on human review of conversations first flagged by Qwen3.5. This design substantially reduces false positives, and the validation study in Table~\ref{tab:qwen-performance} suggests that missed positives are likely limited, given the model's very low false-negative rate.

\begin{figure}[t]
    \centering
    \includegraphics[width=0.85\columnwidth]{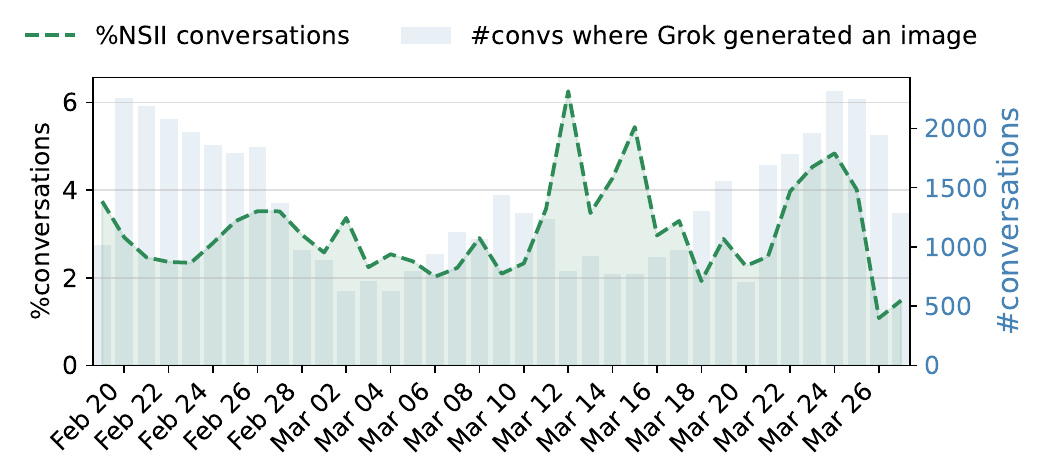}
    \vspace{-4mm}
    \caption{Daily prevalence of NSII conversations.}
    \label{fig:nsii-over-time}
    \vspace{-4mm}
\end{figure}

We also examine the temporal dynamics of NSII conversations to determine whether they occur in a short burst or persist throughout the observation period.
Figure~\ref{fig:nsii-over-time} reports daily prevalence alongside the number of image-generation conversations observed each day. NSII-positive conversations appear throughout the period rather than being confined to a single spike. The daily share fluctuates, with most days falling in the low single digits and several days rising above 4\%, including one peak slightly above 6\%. These temporal patterns suggest that the aggregate estimate is not driven solely by a single anomalous burst of activity; instead, public NSII generation recurs across the observation period, while overall image-generation volume also varies.

\noindent \textbf{User Analysis.} 
We perform a user-level analysis to assess whether NSII generation is concentrated among repeat requesters. The 1,526 confirmed NSII cases involved 839 users, representing 4.2\% of image requesters in our dataset. Although most were associated with only one case, 170 repeat requesters (20.3\%) accounted for 56.2\% of cases, with the nine most active alone accounting for 20.3\%. Requests from repeat requesters contributed a substantial share of daily confirmed NSII cases throughout the observation period. Appendix~\ref{app:user-analysis} provides the detailed user-level and temporal analyses.

\begin{table*}[!t]
\centering
\footnotesize
\setlength{\tabcolsep}{4pt}

\begin{tabular}{
p{0.14\textwidth}
>{\raggedleft\arraybackslash}p{0.06\textwidth}
p{0.32\textwidth}
p{0.40\textwidth}
}
\toprule
\textbf{Concept name} & \textbf{\%convs.} & \textbf{Concept prompt} & \textbf{Example summary} \\
 \toprule
 Increase Revealingness & 60.4\% & Does the request ask to make the subject's clothing more revealing, expose more skin, or increase cleavage/cleavage exposure? & The user requests Grok to modify a reference image by altering the shirt to have the top two buttons open, specifically describing the desired outcome as ``sexy cleavage exposing.''...\\ %
\midrule
Sexualized Outfit Change & 54.1\% & Does the request ask to change the subject's clothing to a more revealing, fetishized, or sexualized outfit (e.g., bunny, latex, maid, bikini)? & The user asks Grok to modify a reference image by changing the subject's outfit to a black bunny suit...\\ %
\midrule
Clothing Removal / Nudity Increase & 42.2\% & Does the request explicitly ask to remove clothing or otherwise increase the amount of exposed skin/body parts? & The user asks Grok to modify a reference image by opening the subject's clothes. The requested action is best summarized as a clothing removal or body exposure increase, specifically asking to unbutton or open the denim shirt worn by the subject in the context image...\\
\midrule
Specific Costume Swap & 28.4\% & Does the request specify swapping the subject's current outfit for a particular sexualized costume or themed attire (e.g., bunny, maid, nurse, latex, fairy)? & The user asks Grok to swap a specific outfit onto a public figure, referencing a prior request to apply a gothic bunny costume to Donald Trump. The requested action is best summarized as a sexualized outfit change, involving the transfer of a revealing bodysuit, thigh-high boots, and bunny ears onto the subject's body...  \\
\midrule
Add Visual Effects / Props & 25.5\% & Does the request add visual effects (e.g., sweat, wetness, water) or accessories/extra characters to the image? & The user asks Grok to modify a reference image of a person in a white tank top by adding a second individual described as a 'cute twink boyfriend' embracing them...\\ %
\midrule
Pose or Angle Change & 20.2\% & Does the request specify a change in the subject's pose, position, or viewing angle (e.g., doggy‑style, bent over, front view, rear view)? & The user issues a follow-up request to Grok, expressing dissatisfaction with a previous output and instructing the AI to reposition the subject into a 'doggy position.'... \\
\midrule
Material/Transparency Change & 14.7\% & Does the request involve changing the material, texture, or transparency of the clothing (e.g., latex, sheer, mesh, semi‑transparent, opaque to transparent)? & The user requests Grok to modify a reference image by changing the dress color to transparent red. The requested action is best characterized as a sexualized outfit change that involves increasing body exposure through transparency, rather than complete clothing removal...\\
\midrule
Body Part Enlargement & 10.3\% & Does the request ask to enlarge or otherwise modify specific body parts (e.g., breasts, buttocks, hips, waist) for a more sexualized appearance? & The user asks Grok to modify a reference image of a woman from behind by increasing the size of her buttocks... \\
\midrule
Garment Structural Edit & 9.9\% & Does the request ask for structural modifications to the garment such as shortening hems, adding slits, splitting a dress, adding zippers, straps, or other cut‑outs? & The user asks Grok to modify an image by altering the subject's clothing, specifically requesting to split the dress into two separate pieces and adjust the fit. The requested transformation involves changing the garment into a two-piece outfit where the top is described as 'more open' and the bottom skirt is made 'shorter'... \\ %
\midrule
Swimwear/Bikini Replacement & 5.2\% & Does the request specifically ask to replace the current clothing with swimwear, bikini, micro‑bikini, or similar beachwear? & The user asks Grok to modify a reference image by changing the subject's outfit into a black micro bikini...\\ %
\midrule
Facial Feature Modification & 5.1\% & Does the request involve altering the subject's facial expression or features, such as opening the mouth, extending the tongue, or creating a grotesque face? & The user requests Grok to generate an image of a woman with her tongue sticking out in a sexual manner, referencing a prior context image...\\ %
\bottomrule
\end{tabular}
\vspace{-2mm}
\caption{Concepts extracted with the LLooM framework that describe the user requests in NSII conversations. We report the concept name, its prevalence in our dataset, the prompt used to determine relevance to the conversation, as well as an example summary of the conversation from Qwen3.5 (truncated for readability purposes).}
\label{tab:rq2_concept_set}
\vspace{-2mm}
\end{table*}

\subsection{RQ2: Manifestation of NSII on X}

Having measured the prevalence of public-facing Grok-powered NSII, we next characterize how this abuse manifests in the wild. Specifically, we ask what kinds of NSII requests users make to Grok on X.
To shed light on this, we use the LLooM framework to extract concepts mentioned in our NSII-positive conversations.
Table~\ref{tab:rq2_concept_set} reports the extracted concepts, their prevalence, as well as the prompt used by LLooM to determine relevance to the concept and an example. Note that a conversation can have multiple concepts (e.g., a user requesting multiple and diverse image edits).
The results in Table~\ref{tab:rq2_concept_set} show that Grok-powered NSII is dominated by requests that sexualize an existing person through image-editing operations. The most common concept is ``Increase Revealingness,'' which appears in 60.4\% of NSII conversations, followed by ``Sexualized Outfit Changes'' (54.1\%) and ``Clothing Removal or Nudity Increase'' (42.2\%). In practice, this means that users often do not need to ask Grok to generate an explicitly sexual image from scratch. Instead, they request concrete visual edits, such as exposing more skin, changing clothing, opening garments, or making the target’s appearance more suggestive. This pattern suggests that NSII on X often takes the form of incremental transformation: ordinary or publicly available images are gradually modified into sexualized outputs.
Also, our concept-based analysis highlights the diversity of strategies through which users operationalize these transformations. Some requests are direct textual instructions: for example, ``Sexualized Outfit Change'' (54.1\%) often appear as prompts asking Grok to place the target in a particular revealing or fetishized outfit like a bunny suit. Other requests can reach a similar end state through a more multimodal editing workflow. ``Specific Costume Swap'' (28.4\%), for instance, typically involves providing one image of the target and another image of a costume or clothing item, then asking Grok to transfer that outfit onto the target. This distinction shows that a harmful request is not always contained in a single explicit phrase; it can emerge from the relationship between the text prompt and the attached images.

Beyond outfit transformations, Table~\ref{tab:rq2_concept_set} captures a wider repertoire of sexualizing edits that operate through the body, clothing, pose, and surrounding scene. Some requests intensify the image by adding contextual cues, such as visual effects or props (25.5\%): for example, adding another person to create an intimate, couple-like pose, or introducing effects like wetness that make an otherwise ordinary image look sexualized. Others work through bodily positioning, as in pose or angle changes (20.2\%), where users ask Grok to rotate the subject, alter the viewing angle, or place the target into a sexually suggestive posture (e.g., opening legs or putting them in doggy position). These requests are important because they show that NSII does not depend only on removing clothing; it can also be produced by changing how a body is framed, positioned, or placed in the image. %

At the same time, we find concepts that involve more granular edits to clothing and body appearance. Material or transparency changes (14.7\%) ask Grok to make clothing sheer, mesh, latex-like, or semi-transparent, creating exposure without explicitly requesting nudity. Body-part enlargement (10.3\%) sexualizes the target by exaggerating specific physical features, while garment-structure edits (9.9\%) modify clothing by shortening hems, adding slits, cut-outs, or more open fits. Even lower-frequency categories, such as swimwear or bikini replacement (5.2\%) and facial-feature modification (5.1\%), point to specialized but meaningful pathways into sexualization. Overall, these concepts show that Grok-enabled NSII is a diverse and compositional editing practice: users can sexualize a real person through many small transformations that may look innocuous in isolation. However, they become harmful when applied to a real target in a multimodal image-editing context. This has direct implications for mitigation, as safeguards need to reason over the full interaction, including the target image, reference images, requested edits, and cumulative sexualized effect.

\begin{figure}[t]
    \centering
    \includegraphics[width=0.8\columnwidth]{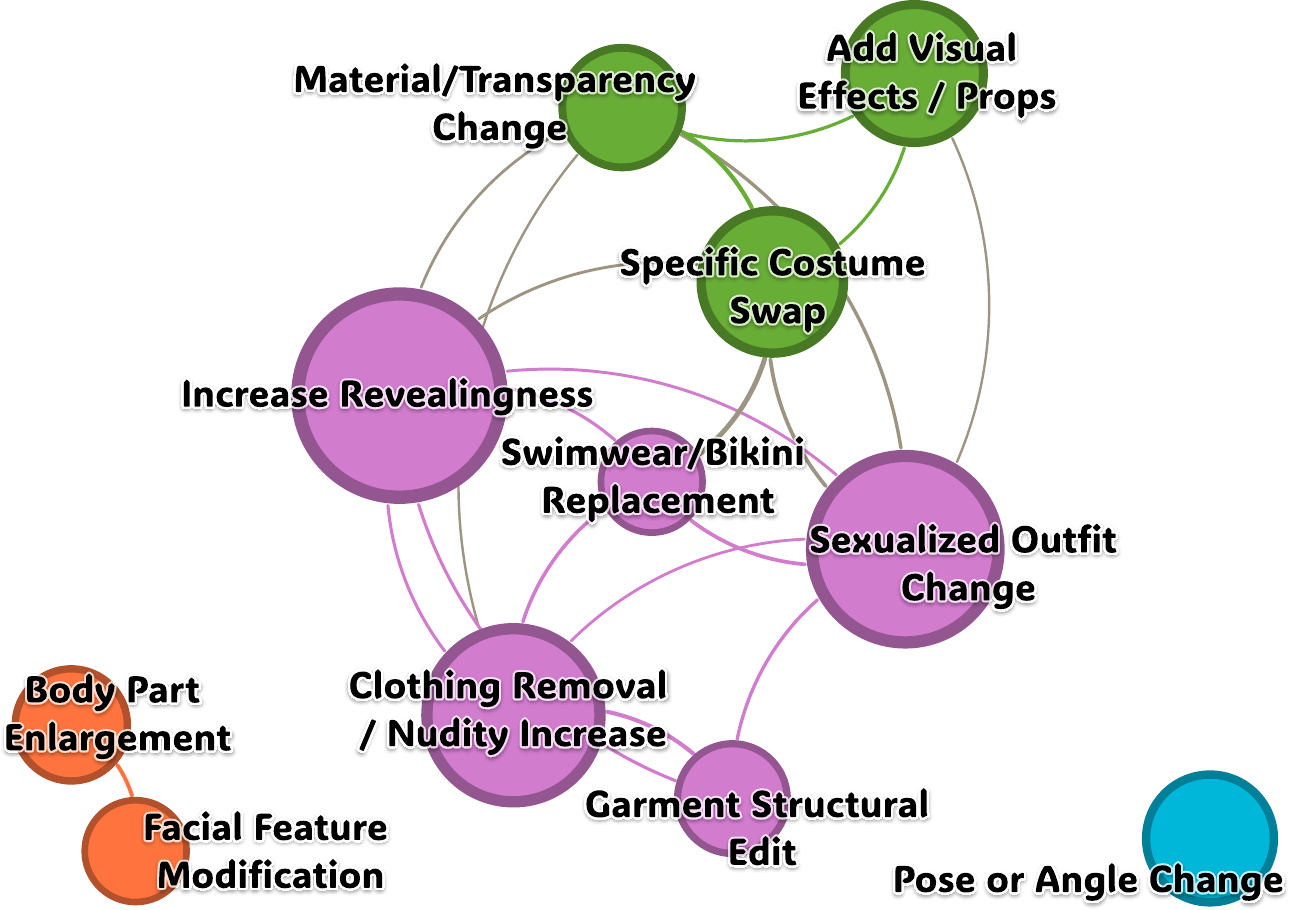}
    \caption{Significant co-occurrence network among the 11 LLooM-derived request-type concepts. Edges are shown only for concept pairs whose overlap exceeds chance expectation and node colors indicate Leiden communities.}
    \label{fig:rq2-concept-network}
    \vspace{-4mm}
\end{figure}

\smallskip \noindent \textbf{Concept Co-occurrence.} As mentioned above, an NSII conversation may include multiple requests for image modifications; therefore, here we aim to study this phenomenon and the types of requests that co-appear together.
Our results show that 21\% of the NSII consist of only one user request, while 75\% of the NSII conversations involve 2 or more concepts. Also, we find that 4\% of the conversations are not captured by the list of eleven concepts, highlighting that our list of concepts largely cover our dataset of NSII conversations.
Overall, these results suggest that Grok-powered NSII is often comprising several transformation strategies. This motivates the co-occurrence analysis below, which examines which concepts tend to appear together and whether these combinations form coherent families of request behavior.

To examine these co-occurrence patterns, we construct a concept network in which edge weights reflect the association strength $w_{ij} = N * c_{ij} / (c_i c_j)$ between concept pairs~\cite{ech2009conceptcooccurance}, where $N$ is the total number of conversations, $c_{ij}$ is the number of conversations labeled with both concepts $i$ and $j$, and $c_i$ and $c_j$ are the frequencies of concepts $i$ and $j$. We retain only edges whose overlap exceeds chance under a one-sided hypergeometric test with Benjamini-Hochberg correction ($q < 0.05$), and perform community detection using the Leiden algorithm~\cite{traag2019louvain}. Figure~\ref{fig:rq2-concept-network} shows that NSII requests to Grok organize into four communities. 
The largest community (purple) captures the main clothing-and-exposure pathway into Grok-powered NSII. It groups ``Increase Revealingness,'' ``Sexualized Outfit Change,'' ``Clothing Removal/Nudity Increase,'' ``Swimwear/Bikini Replacement,'' and ``Garment Structural Edit.'' This community indicates that requests to sexualize clothing, expose more skin, remove or alter garments, and replace outfits with swimwear tend to appear together. In practice, this suggests that many harmful Grok interactions are organized around incremental modification of a target’s clothing and degree of bodily exposure. The green community captures a related but distinct set of contextual editing strategies. It includes ``Specific Costume Swap,'' ``Add Visual Effects/Props,'' and ``Material/Transparency Change.'' These concepts are connected to the purple community, but form their own community, suggesting that they operate as a separate pathway into sexualization. Rather than simply asking for more revealing clothing, users may provide a costume reference, request a material change such as transparency, or add scene elements that make the image more sexualized. 
The smaller orange community links ``Body Part Enlargement'' and ``Facial Feature Modification,'' indicating a more targeted form of sexualization focused on altering body/face attributes. Finally, ``Pose or Angle Change'' appears as a separate community, suggesting that pose-based requests do not strongly co-occur with the other concepts.

\subsection{RQ3: Targets of NSII on X}

We next characterize the targets of Grok-generated NSII, focusing on how many people are targeted in each conversation and what attributes can be inferred about them.
Our analysis shows that Grok-generated NSII in our dataset is overwhelmingly directed at a single target per conversation. Specifically, we find that 85.9\% of NSII-positive conversations contain one target, while 11.6\% contain two targets. Conversations involving three or more targets are less frequent, together accounting for only 2.6\% of cases. This distribution suggests that NSII use of Grok is primarily individualized: users most often invoke the system to sexualize a specific person rather than to generate multi-target scenes. At the same time, the non-trivial share of 2+ target conversations indicates that some requests ask for multiple people to be jointly placed into a sexualized context.

Having established that most conversations center on a single person, we next examine the perceived gender presentation of the targets. We use perceived gender presentation to describe how a target is visually or contextually presented in the generated output, rather than to infer or assert the target's gender identity. In practice, this captures whether the target is presented in a feminine, masculine, androgynous, or unclear way based on the available image and conversation context. 
We find that the distribution is highly imbalanced: 80.9\% of the target entries are coded as feminine-presenting, while 19.1\% are coded as masculine-presenting. 
No target entries are coded as androgynous or unclear. 
This pattern shows that Grok-generated NSII in our dataset disproportionately targets feminine-presenting people, while also indicating that masculine-presenting targets account for a meaningful minority of cases.

\begin{figure}[t]
    \centering
    \includegraphics[width=0.85\columnwidth]{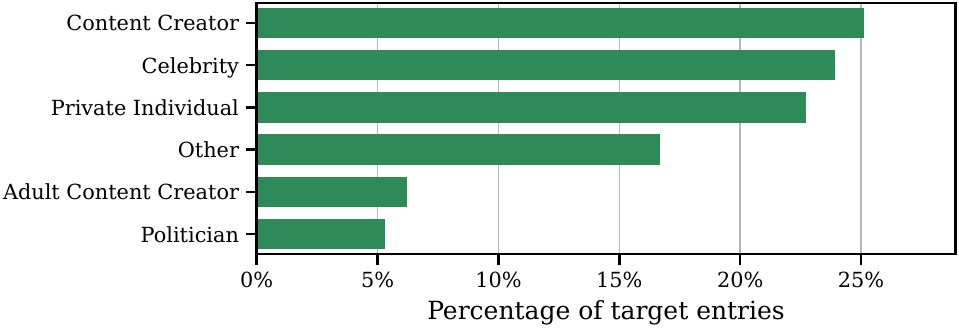}
    \vspace{-3mm}
    \caption{Public role status of the NSII targets.}
    \label{fig:nsii-public-status}
    \vspace{-4mm}
\end{figure}

We then examine what kinds of public roles these targets occupy. 
Figure~\ref{fig:nsii-public-status} shows that the targets are distributed across several public-role categories. The largest category is content creators, with 458 target entries (25.1\%), followed closely by celebrities with 436 entries (23.9\%) and private individuals with 414 entries (22.7\%). Other targets account for 304 entries (16.7\%), while adult content creators and politicians account for smaller shares, with 113 entries (6.2\%) and 97 entries (5.3\%), respectively.
Our results show that Grok-generated NSII is not limited to one type of target. Although highly visible figures such as celebrities, politicians, and content creators are common, nearly one quarter of target entries are private individuals. This finding is suggests that public visibility is not a necessary condition for being targeted: Grok's image-generation and editing capabilities can also be used against people who do not have celebrity or political roles.

\begin{figure}[t]
    \centering
    \includegraphics[width=0.85\columnwidth]{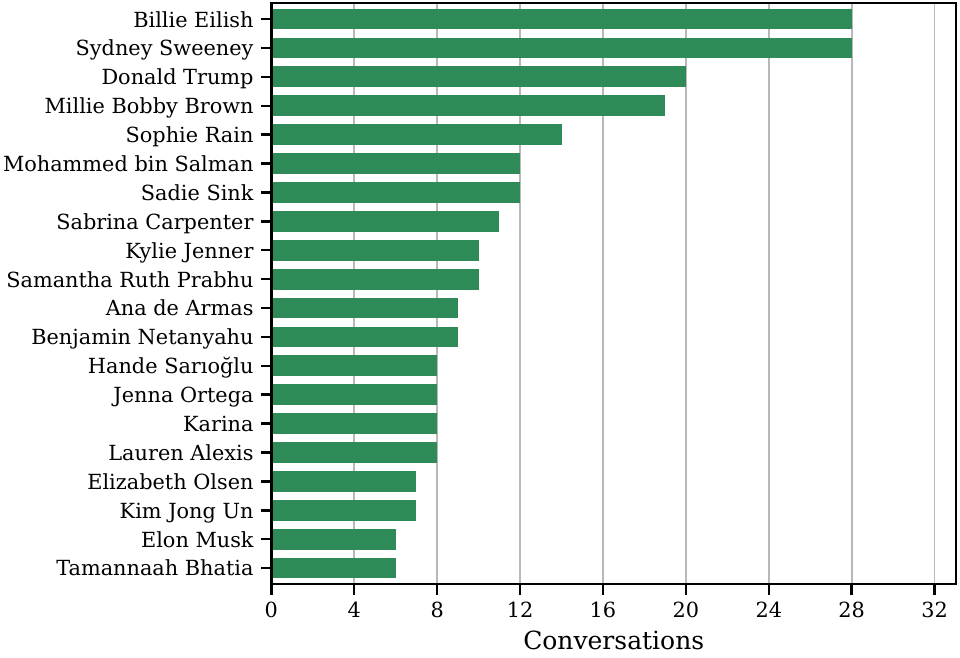}
    \vspace{-3mm}
    \caption{Top targets in our dataset.}
    \label{fig:nsii-top-targets}
    \vspace{-4mm}
\end{figure}

Finally, Figure~\ref{fig:nsii-top-targets} examines the most frequently identified named targets. The counts are relatively dispersed: the two most frequent named targets, Billie Eilish and Sydney Sweeney, each appear in 28 conversations, followed by Donald Trump in 20 conversations and Millie Bobby Brown in 19. This indicates that NSII targeting spans a broad set of recognizable public figures. Importantly, the list underscores the global nature of Grok-enabled NSII targeting. The most frequently identified targets include U.S. entertainers and influencers, Indian actors, a Turkish media figure, a Korean pop artist, technology and business figures, and political leaders from the United States, Saudi Arabia, Israel, and North Korea.
Ironically, Elon Musk also appears among the top identified targets, despite being the owner of both X and xAI, the company behind Grok.
Overall, while many named targets are highly visible figures, the pattern is not confined to one national context, profession, or media ecosystem. Grok-generated NSII appears to draw on a transnational pool of recognizable people, reflecting how public interactions with a globally available generative system can reproduce sexualized targeting across the entire globe.

Taken together, these target-level results show that Grok-generated NSII is not confined to isolated abusive requests against a small set of highly visible figures. Instead, it appears as a recurrent pattern of platform-enabled targeting: it is predominantly individualized, disproportionately directed at feminine-presenting people, spans both public figures and private individuals, and draws from a transnational pool of recognizable targets. This pattern is directly relevant to the systemic-risk framework of the Digital Services Act, under which very large online platforms are expected to assess and mitigate risks arising from the design, functioning, and use of their services, including risks related to illegal content, fundamental rights, gender-based violence, and serious harms to physical and mental well-being~\cite{EU_DSA_2022}. 
Our work provides empirical evidence and emphasizes that Grok-enabled NSII should be understood as a platform-level risk, not merely as a collection of disconnected user abuses.

\section{Ethical Considerations}
\label{sec:ethics}

Our work was approved  by our institution's Ethical Review Board. Also, we emphasize that our work relies on publicly available data shared on X.
To safeguard user privacy, we anonymize user names and identifiers before storing the data on our servers. Also, we store the data in secure servers within our organization's infrastructure, and we do not share any data with third parties. 
For data analysis, we used only open-source models run locally on our infrastructure; we do not reveal any sensitive data or images of any depicted individual in our dataset.

Our data collection relied on researcher-controlled sockpuppet accounts that repeatedly accessed publicly visible Grok interactions on X. This approach likely did not comply with X’s platform rules.
However, we view this as a last-resort method that was used only after pursuing official access channels. Before collecting data from X using the approach described in this paper, we first sought access through the platform’s official research access mechanisms under Article 40(12) of the Digital Services Act, which provides a pathway for researchers to request access to platform data that is publicly available on the platform for the study of systemic risks. In late 2025, our research team submitted a data access request to study systemic risks associated with Grok on X, but the platform declined to provide access.

Our work also carries a potential misuse risk: by identifying the types of requests that resulted in Grok generating NSII, the paper could inadvertently reveal which kinds of abusive image-editing instructions were successful. In other words, our analysis may indicate what has worked in practice for users attempting to produce such content. We mitigate this risk by reporting request types only at an aggregate and normalized level. We do not include raw conversations, generated images, or detailed prompt examples that would enable replication of abusive requests.

Nevertheless, we argue that the public-interest benefits of our research outweigh the potential limited harm introduced by our data collection and results. NSII is an important form of technology-facilitated abuse, and studying it is necessary to understand how AI systems may enable or amplify sexually related harm in practice. Examining Grok's role in these interactions also serves an accountability function, especially in a setting where the platform denied official data access. More broadly, independent empirical scrutiny is necessary to document emerging systemic risks and inform platform governance and policymakers.

\section{Discussion \& Conclusion}

This paper provides an empirical analysis of Grok-enabled NSII on X. Using a large-scale sample of public Grok interactions, we identify approximately 1.5K confirmed conversations in which users explicitly requested NSII and Grok complied, corresponding to 2.94\%--3.05\% of analyzable image-generation conversations. We show that this abuse is recurring over time, often operates through incremental image-editing requests, and targets a diverse set of people, with a strong concentration on feminine-presenting individuals. Our findings show that public, platform-integrated image generation can create systemic risks that require stronger safeguards, effective reporting mechanisms, and independent oversight.
We discuss the implications below.

\smallskip \noindent \textbf{Grok-Enabled NSII as a Systemic Risk.} Our prevalence results show that Grok-enabled NSII is a recurring form of public image generation on X. Across 50K conversations in which Grok generated an image, we identified $\approx1.5K$ confirmed cases where a user explicitly requested NSII and Grok complied, corresponding to 2.94\%--3.05\% of image-generation conversations. 
At first glance, these rates may appear low. We argue, however, that these prevalence estimates are significant and support characterizing this emerging form of abuse as a systemic risk. Three aspects support this interpretation.
First, even if NSII represents a minority of image-generation activity, each case involves the public production of sexualized synthetic imagery of a real person and can be immediately viewed, shared, or further amplified. So the harm and impact to individuals can be substantial.
Second, the daily prevalence pattern shows that confirmed NSII conversations appear across the observation period rather than being confined to a single anomalous spike, indicating the recurring nature of this abuse phenomenon.
Third, Grok-enabled NSII targets a diverse set of individuals, including celebrities across the globe, politicians, content creators, adult content creators, and private individuals. This diversity suggests that the abuse is not concentrated around a small number of celebrities. Rather, it can affect different kinds of people across different social and geographic contexts.

Our findings suggest that Grok-enabled NSII should be treated as a recurring platform-level risk arising from the service's design and use, rather than reduced to isolated user violations. This has direct implications for platform governance. If NSII is recurring and spans many types of targets and platforms, it should not be treated as isolated user misconduct. Instead, platforms should continuously monitor public AI image generation, implement safeguards before harmful outputs are posted, and provide dedicated, effective reporting mechanisms for affected individuals.

\smallskip \noindent \textbf{Beyond Nudification: NSII as Incremental Image Editing.}
Our manifestation results show that Grok-enabled NSII extends beyond direct requests for nudification or explicit sexual imagery. The most common request types involve increasing revealingness, changing clothing into sexualized outfits, and removing or opening clothing. Other recurring concepts include costume swaps, material or transparency changes, pose or angle changes, body-part modification, and the addition of props or visual effects. These results suggest that much of the abuse operates through incremental image editing.
This has important implications for content moderation. Moderation approaches that focus narrowly on explicit sexual keywords or nudity detection are unlikely to capture the full range of harmful behavior documented in our study. A request to change clothing, modify pose, or transfer a costume onto a target may not appear abusive when viewed in isolation. The harmfulness often emerges from the combination of the target image, the reference image, the requested edit, and the sexualized effect of the output.

Safeguards for platform-integrated image generation therefore need to reason over the full conversation and multimodal interaction. This includes the context tweet, the user's invocation, any attached or referenced media, and the generated response. 
In practice, we argue that it is important that the platforms implement pre-publication checks that assess whether a requested edit sexualizes a real person and design effective content moderation systems that take into account information at the conversation level so that context, source imagery, and generated media are considered together.

\smallskip \noindent \textbf{Gendered and Individualized Targeting.}
Most confirmed NSII conversations in our dataset target a single person, and most target entries are feminine-presenting.
The abuse is therefore individualized and strongly gendered.
The target pool also extends beyond celebrities. Although content creators, celebrities, politicians, and other public-facing figures appear frequently, private individuals account for nearly one quarter of target entries. This finding emphasizes that public visibility is not a necessary condition for being targeted. People without celebrity status or political prominence may still become targets when their images are available in public online contexts.
This has practical implications for victim protection. Platforms should design NSII interventions that cover both public figures and private individuals, rather than focusing primarily on celebrities. Detection, reporting, and removal systems should account for users who may have fewer resources to identify harmful content, request takedown, or prevent further circulation. This is especially important because prior audit work on X found that reporting mechanisms for non-consensual intimate media can be highly ineffective: \citet{qiwei2024reporting} found that images reported through X's non-consensual nudity pathway were not removed over a three-week period, while images reported about copyright issues were promptly removed. Effective remedies should therefore include dedicated reporting pathways for AI-generated sexualized imagery, rapid removal procedures, and mechanisms for identifying repeated targeting of the same person.

\smallskip \noindent \textbf{Limitations \& Future Work.}
Our study has some limitations. First, our prevalence estimate applies to the public Grok conversations we collected and to the subset of image-generation conversations whose media remained retrievable at annotation time. We could not analyze 1,436 conversations for which the media were no longer accessible.
Second, our final confirmed NSII set was constructed by manually reviewing conversations flagged by the automated detector, which is likely to miss some instances of NSII. This review relied on a single annotator, whose judgments may differ from those of other annotators. Missed cases could lead to underestimation of prevalence, while incorrectly included cases could lead to overestimation. Errors that vary across request types or target categories could also bias downstream comparisons. Our validation does not quantify how annotation and classification uncertainty propagates to these results, as this will require more researchers exposed to potentially harmful content. 
Third, our analysis focuses on public interactions with Grok on X. It does not measure private Grok use, deleted conversations, downstream resharing, or victim experiences after publication. 
Fourth, our data collection window (February 19--March 27, 2026) falls entirely after the mitigation measures announced by X on January 9 and January 14, 2026. These measures restricted image generation and editing through the Grok account to paid subscribers~\cite{grok_restriction_paid} and introduced technical restrictions intended to prevent the editing of images of real people into revealing clothing~\cite{grok_safety_updates}. Accordingly, the observed prevalence should be interpreted as a post-mitigation rate for public Grok conversations rather than as a general baseline for Grok use. Because our dataset does not include a comparable pre-mitigation period or a longer observation window, we cannot determine whether the prevalence of NSII increased, decreased, or remained stable over time.
Fifth, whether imagery was intimate was left to the annotators' interpretation, introducing subjectivity and potentially limiting coverage of culturally or contextually specific forms of intimacy.
Finally, our target annotation for public status achieved 52\% accuracy, with errors involving misclassification between private individuals, content creators. We therefore interpret the estimated share of private individuals as conservative, since some targets assigned to other categories may also be private individuals.
Future work should examine how public AI-generated NSII spreads after creation, how victims encounter and report such content, and how platform interventions affect recurrence over time. Comparative studies across different platform-integrated AI systems would also help determine whether the patterns observed here are specific to Grok on X or reflect a broader class of risks.

\section*{Acknowledgments} 
This work is part of the project ``EGOS: Effective Governance for cybersecurity and Online Safety'' with file number KICH1.VE05.23.001 of the research programme ``Cybersecurity Research Program 2023,'' which is financed by the Dutch Research Council (NWO) under the grant ID https://doi.org/10.61686/GXRYB36907. 
Also, this work was partially funded by an unrestricted gift from Google.
Savvas Zannettou was supported by the European Commission's Joint
Research Centre (JRC) under Expert Contract No.~CT-EX2023D894031-101.
Claude Code and OpenAI's Codex were used as coding assistants (to write code to generate plots and develop dashboards for annotations). We also used ChatGPT for fixing grammar and writing issues in the paper. The authors take full responsibility for the content, analyses, and conclusions presented in this work.

\bibliography{aaai2026}

\clearpage
\normalsize
\section*{Paper Checklist}

\begin{enumerate}

\item For most authors...
\begin{enumerate}
    \item  Would answering this research question advance science without violating social contracts, such as violating privacy norms, perpetuating unfair profiling, exacerbating the socio-economic divide, or implying disrespect to societies or cultures?
    \answerYes{Yes.}
  \item Do your main claims in the abstract and introduction accurately reflect the paper's contributions and scope?
    \answerYes{Yes.}
   \item Do you clarify how the proposed methodological approach is appropriate for the claims made? 
    \answerYes{Yes, in ``Methodology''.}
   \item Do you clarify what are possible artifacts in the data used, given population-specific distributions?
    \answerYes{Yes, we discuss our data limitations in ``Discussion and Conclusion'' section.}
  \item Did you describe the limitations of your work?
    \answerYes{Yes, the limitations are discussed in ``Discussion \& Conclusion''}
  \item Did you discuss any potential negative societal impacts of your work?
    \answerYes{Yes, in ``Ethical Considerations.''}
      \item Did you discuss any potential misuse of your work?
    \answerYes{Yes, in ``Ethical Considerations.''}
    \item Did you describe steps taken to prevent or mitigate potential negative outcomes of the research, such as data and model documentation, data anonymization, responsible release, access control, and the reproducibility of findings?
    \answerYes{Yes, in ``Ethical Considerations.''}
  \item Have you read the ethics review guidelines and ensured that your paper conforms to them?
    \answerYes{Yes.}
\end{enumerate}

\item Additionally, if your study involves hypotheses testing...
\begin{enumerate}
  \item Did you clearly state the assumptions underlying all theoretical results?
    \answerNA{NA}
  \item Have you provided justifications for all theoretical results?
    \answerNA{NA}
  \item Did you discuss competing hypotheses or theories that might challenge or complement your theoretical results?
    \answerNA{NA.}
  \item Have you considered alternative mechanisms or explanations that might account for the same outcomes observed in your study?
    \answerNA{NA}
  \item Did you address potential biases or limitations in your theoretical framework?
    \answerNA{NA}
  \item Have you related your theoretical results to the existing literature in social science?
    \answerNA{NA}
  \item Did you discuss the implications of your theoretical results for policy, practice, or further research in the social science domain?
    \answerNA{NA}
\end{enumerate}

\item Additionally, if you are including theoretical proofs...
\begin{enumerate}
  \item Did you state the full set of assumptions of all theoretical results?
    \answerNA{NA.}
	\item Did you include complete proofs of all theoretical results?
    \answerNA{NA.}
\end{enumerate}

\item Additionally, if you ran machine learning experiments...
\begin{enumerate}
  \item Did you include the code, data, and instructions needed to reproduce the main experimental results (either in the supplemental material or as a URL)?
    \answerYes{Yes. The prompts are shared in the online repository.}
  \item Did you specify all the training details (e.g., data splits, hyperparameters, how they were chosen)?
    \answerNA{NA}
     \item Did you report error bars (e.g., with respect to the random seed after running experiments multiple times)?
    \answerNA{NA}
	\item Did you include the total amount of compute and the type of resources used (e.g., type of GPUs, internal cluster, or cloud provider)?
    \answerYes{Yes, we use models run locally on 2 H200NVL NVIDIA GPUs using vLLM (see Appendix).}
     \item Do you justify how the proposed evaluation is sufficient and appropriate to the claims made? 
    \answerYes{Yes, see  ``Methodology'' section.}
     \item Do you discuss what is ``the cost`` of misclassification and fault (in)tolerance?
    \answerNA{NA.}
  
\end{enumerate}

\item Additionally, if you are using existing assets (e.g., code, data, models) or curating/releasing new assets, \textbf{without compromising anonymity}...
\begin{enumerate}
  \item If your work uses existing assets, did you cite the creators?
    \answerNA{NA}
  \item Did you mention the license of the assets?
    \answerNA{NA}
  \item Did you include any new assets in the supplemental material or as a URL?
    \answerNA{NA}
  \item Did you discuss whether and how consent was obtained from people whose data you're using/curating?
    \answerNA{NA.}
  \item Did you discuss whether the data you are using/curating contains personally identifiable information or offensive content?
\answerNA{NA.}
\item If you are curating or releasing new datasets, did you discuss how you intend to make your datasets FAIR?
\answerNA{NA.}
\item If you are curating or releasing new datasets, did you create a Datasheet for the Dataset? 
\answerNA{NA.}
\end{enumerate}

\item Additionally, if you used crowdsourcing or conducted research with human subjects, \textbf{without compromising anonymity}...
\begin{enumerate}
  \item Did you include the full text of instructions given to participants and screenshots?
    \answerNA{NA.}
  \item Did you describe any potential participant risks, with mentions of Institutional Review Board (IRB) approvals?
    \answerNA{NA.}
  \item Did you include the estimated hourly wage paid to participants and the total amount spent on participant compensation?
    \answerNA{NA.}
   \item Did you discuss how data is stored, shared, and de-identified?
   \answerYes{Yes, see ``Ethical Considerations'' section.}
\end{enumerate}

\end{enumerate}

\appendix

\begin{table*}[h]
\small
\centering
\begin{tabular}{@{}p{4.0cm}p{5.0cm}p{6.6cm}@{}}
\toprule
\textbf{Stage} & \textbf{Parameter} & \textbf{Value} \\
\midrule
Concept generation & Embedding model & \texttt{codefuse-ai/F2LLM-4B} \\
Concept generation & LLM & \texttt{openai/gpt-oss-120b} \\
Concept generation & Decoding & Temperature \(=0.0\), top-\(p=0.9\) \\
Distill & Max tokens & \(6000\) \\
Synthesize & Max tokens & \(6000\) \\
\midrule
Concept scoring & Scoring model & \texttt{openai/gpt-oss-120b} \\
Concept scoring & Decoding & Temperature \(=0.0\), top-\(p=1.0\) \\
Score & Max tokens & \(300\) \\
\bottomrule
\end{tabular}
\caption{Models and parameters used in LLooM.}
\label{tab:lloom_hyperparams}
\end{table*}

\section{Implementation Details}

\subsection{Infrastructure}
We deployed Qwen3.5-122B-10B in our local infrastructure, which consists of 2H200 NVL NVIDIA GPUs. For inference, we use the vLLM library to serve the model.

\subsection{LLooM Hyperparameters}
\label{app:lloom-hyperparameters}

Table~\ref{tab:lloom_hyperparams} reports the LLooM model and decoding configuration used for concept generation and scoring in this study.

\section{User-Level Analysis}
\label{app:user-analysis}

\begin{figure}[t]
    \centering
    \includegraphics[width=0.9\linewidth]{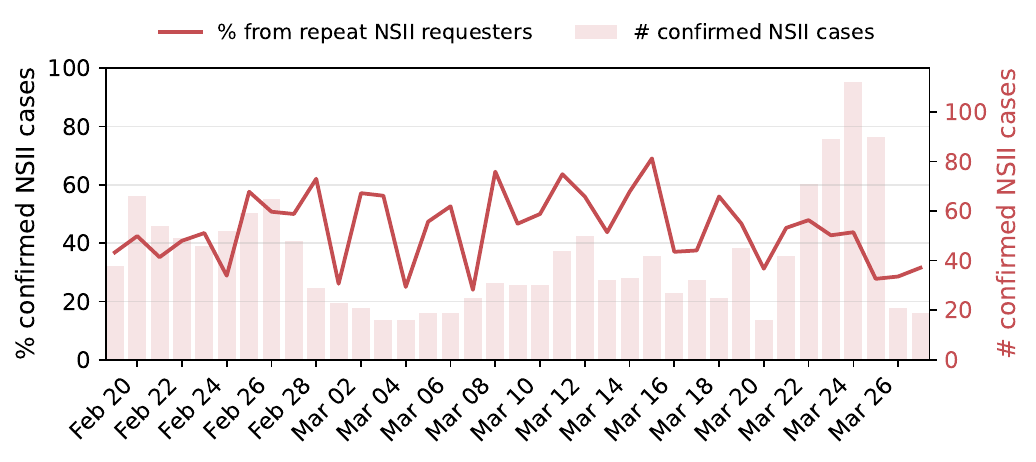}
    \caption{Daily contribution from users with previously confirmed NSII requests.}
    \label{fig:nsii-repeat-requests}
\end{figure}

To assess whether NSII generation reflects broad participation or is disproportionately driven by a small group of persistent users, we perform a user-level analysis.
Among the 19,998 distinct users who made a request to Grok to generate an image in our dataset, 839 (4.2\%) are associated with at least one of the 1,526 human-confirmed NSII cases. We define a one-off requester as a user associated with exactly one confirmed NSII case, and a repeat requester as a user associated with at least two confirmed NSII cases.
Most NSII requesters were one-off requesters (669/839; 79.7\%), while the 170 repeat requesters (20.3\%) collectively accounted for 857/1,526 cases (56.2\%), including their initial cases. 
We also find that the nine most active NSII requesters (approximately the top 1\% of all users that requested NSII) accounted for 309 cases (20.3\%).
Figure~\ref{fig:nsii-repeat-requests} shows how the daily share of repeat NSII requests varied over the observation period. We find that repeat requests contribute a substantial share of confirmed NSII generation throughout the observation period, rather than being confined to an isolated spike.
Overall, our user-level analysis shows that NSII generation spans a broad user base but is disproportionately sustained by a smaller group of repeat requesters, suggesting that targeted restrictions could mitigate (but not eliminate) the problem.

\section{Robustness to the Concept Relevance Threshold}
\label{app:threshold-sensitivity}

LLooM scores each conversation--concept pair on a five-point scale ranging from \emph{strongly agree} (A; score $=1.0$) through \emph{agree} (B; $0.75$), \emph{neutral} (C; $0.5$), and \emph{disagree} (D; $0.25$) to \emph{strongly disagree} (E; $0.0$). In the main analysis, a conversation matches a concept if its score exceeds $0.5$, that is, if the scoring model answers A or B. To test whether our findings depend on including the weaker \emph{agree} label, we repeat the analysis under a stricter rule in which only A counts as a match.

The stricter rule removes 283 of the 4,493 positive conversation--concept labels (6.3\%); by construction, no negative label becomes positive. Table~\ref{tab:threshold_sensitivity} reports the prevalence of each concept under the stricter rule and its relative change with respect to the main analysis. Prevalence decreases for every concept, but only modestly: the median relative change is $-5.1\%$, ranging from $-1.9\%$ to $-11.5\%$. The ordering of concepts is nearly identical under the two rules. The two rankings are strongly correlated (Spearman $\rho=0.991$, 95\% bootstrap CI over conversations $[0.986, 1.000]$, 2,000 resamples; Kendall $\tau=0.964$, CI $[0.944, 1.000]$), and the only change is a swap between the two least prevalent concepts, \emph{Facial Feature Modification} and \emph{Swimwear/Bikini Replacement}, which differ by at most two conversations under either rule.

The reduction is also not concentrated among common or rare concepts. Although the absolute decrease grows with a concept's original prevalence (Spearman $\rho=0.864$, $p<0.001$), which is expected because more prevalent concepts have more positive labels to lose, the relative change is not significantly correlated with original prevalence (Spearman $\rho=0.282$, $p=0.40$). The stricter rule therefore rescales prevalence roughly uniformly rather than favoring particular request types.

The downstream analyses are similarly stable. The mean number of concepts per conversation decreases from 2.94 to 2.76, while the median remains 3. In the co-occurrence network, association-strength lift is highly consistent across all 55 concept pairs (Spearman $\rho=0.982$), and 18 of the 20 significant edges are retained, with no new edges added.

\begin{table}[t]
\centering
\small
\setlength{\tabcolsep}{4pt}
\caption{Concept prevalence among the 1,526 NSII-positive conversations when only \emph{strongly agree} (A) counts as a match, and relative change with respect to the main analysis (A or B). Concepts are ordered by prevalence in the main analysis.}
\label{tab:threshold_sensitivity}
\begin{tabular}{@{}lrr@{}}
\toprule
\textbf{Concept name} & \textbf{Prev. (\%)} & \textbf{Change (\%)} \\
\midrule
Increase Revealingness             & 60.4 & $-7.9$  \\
Sexualized Outfit Change           & 54.1 & $-2.8$  \\
Clothing Removal / Nudity Increase & 42.2 & $-11.5$ \\
Specific Costume Swap              & 28.4 & $-3.6$  \\
Add Visual Effects / Props         & 25.5 & $-5.1$  \\
Pose or Angle Change               & 20.2 & $-3.1$  \\
Material/Transparency Change       & 14.7 & $-8.2$  \\
Body Part Enlargement              & 10.3 & $-10.8$ \\
Garment Structural Edit            &  9.9 & $-1.9$  \\
Facial Feature Modification        &  5.1 & $-6.0$  \\
Swimwear/Bikini Replacement        &  5.2 & $-2.5$  \\
\bottomrule
\end{tabular}
\end{table}

\section{Stability of Leiden Communities}
\label{app:leiden-stability}

We assess whether the concept communities reported in the RQ2 analysis depend on the configuration of the Leiden algorithm~\cite{traag2019louvain}. Our reference configuration applies Leiden to the significant co-occurrence network (20 edges) using the modularity-based RB configuration quality function with resolution $\gamma=1$, association-strength lift as the edge weight, two iterations, and a fixed random seed. It yields four communities: (i) \emph{Increase Revealingness}, \emph{Clothing Removal / Nudity Increase}, \emph{Garment Structural Edit}, \emph{Sexualized Outfit Change}, and \emph{Swimwear/Bikini Replacement}; (ii) \emph{Specific Costume Swap}, \emph{Add Visual Effects / Props}, and \emph{Material/Transparency Change}; (iii) \emph{Body Part Enlargement} and \emph{Facial Feature Modification}; and (iv) \emph{Pose or Angle Change}. We vary the main hyperparameters of the algorithm one at a time and compare each resulting partition to the reference using the adjusted Rand index (ARI). Table~\ref{tab:leiden_stability} summarizes the results.

\textbf{Optimization settings.} The partition is identical across 1,000 random seeds and for 1, 2, 5, or unlimited iterations until convergence.

\textbf{Edge weights.} Replacing association-strength lift with NPMI or raw co-occurrence counts yields the same partition.

\textbf{Resolution.} The reference partition is recovered for $\gamma \in [0.85, 1.25]$. For $0.05 \le \gamma < 0.85$, communities (i) and (ii) merge into a single community, whereas for $1.25 < \gamma \le 1.75$, \emph{Sexualized Outfit Change} and \emph{Swimwear/Bikini Replacement} separate from community (i); in both cases, the remaining groupings are unchanged. The core groupings begin to fragment only at $\gamma \ge 1.8$.

\textbf{Sampling variability.} We further draw 1,000 bootstrap resamples of the 1,526 conversations and rerun the analysis, including edge testing and Leiden. The reference partition is recovered in 66.0\% of resamples (mean ARI $=0.97$); in a further 31.7\%, the only difference is that community (iii) splits into two singletons. The concept triples \{\emph{Increase Revealingness}, \emph{Clothing Removal / Nudity Increase}, \emph{Garment Structural Edit}\} and \{\emph{Specific Costume Swap}, \emph{Add Visual Effects / Props}, \emph{Material/Transparency Change}\} are co-assigned in every resample, and \emph{Pose or Angle Change} is always isolated.

Overall, the community structure is stable across optimization settings, edge weights, and sampling variability. Plausible alternative resolutions merge or split communities around the same core groupings, so the substantive conclusions drawn from the network are unchanged.

\begin{table}[t]
\centering
\small
\setlength{\tabcolsep}{4pt}
\caption{Leiden partitions under alternative configurations, compared with the reference partition (four communities). Rows vary one setting at a time.}
\label{tab:leiden_stability}
\begin{tabular}{@{}llrr@{}}
\toprule
\textbf{Parameter} & \textbf{Setting} & \textbf{\#Comm.} & \textbf{ARI} \\
\midrule
Seed            & 1,000 seeds                   & 4   & 1.00 \\
Iterations      & 1, 2, 5, until convergence    & 4   & 1.00 \\
               
Edge weight     & NPMI, raw co-occurrence       & 4   & 1.00 \\
Resolution      & $\gamma \in [0.85, 1.25]$     & 4   & 1.00 \\
                & $\gamma = 0.5$                & 3   & 0.47 \\
                & $\gamma = 1.5$                & 6   & 0.60 \\
Bootstrap       & 1,000 resamples (mean)        & 4--5 & 0.97 \\
\bottomrule
\end{tabular}
\end{table}

\end{document}